\documentclass[10pt]{iopart}
\usepackage{xcolor}
\usepackage{graphicx}
\usepackage{chemformula}
\usepackage{cite}
\usepackage{siunitx}

\newcommand\hl[1]{%
  \bgroup
  \hskip0pt\color{blue!80!blue}%
  #1%
  \egroup}

\begin{document}

\title[Quantum Heat Under the Microscope: A Perspective on Cryogenic Scanning Thermal Microscopy]{Quantum Heat Under the Microscope: A Perspective on Cryogenic Scanning Thermal Microscopy}

\author{Valentin Fonck${}^\text{1}$, Jean Spiece${}^1$, Pascal Gehring${}^\text{*,1}$}

\address{${}^1$ Institute of Condensed Matter and Nanosciences, Université catholique de Louvain (UCLouvain), 1348 Louvain-la-Neuve, Belgium}
\ead{pascal.gehring@uclouvain.be}
\vspace{10pt}
\begin{indented}
\item[June 2025]
\end{indented}

\begin{abstract}
Exploring thermal transport at cryogenic temperatures presents both significant challenges and valuable insights. By uncovering the thermal counterpart of well-known quantum phenomena, researchers investigated fascinating phenomena ranging from the violation of the Wiedemann-Franz law to the quantisation of phonons. One key frontier remains : no existing method can image local heat transport at the nanoscale under cryogenic conditions. In this Perspective, we review the current state state of the art of local heat transport characterisation techniques and highlight their limitations. As a motivation for the development of cryogenic Scanning Thermal Microscopy, we provide five case studies illustrating how this approach could deepen our understanding of exotic quantum phases and enable the emergence of transformative technologies.

\end{abstract}

%
%
\submitto{Nano Futures}
%
%
\ioptwocol

\section{The need for cryogenic nanoscale thermal characterisation}
\label{sec:intro}

Understanding heat transport at the nanoscale is a central challenge in modern solid-state physics. At low temperatures, classical models of thermal conduction fail, giving way to a rich landscape of quantum regimes. While emerging quantum phenomena in these systems are extensively probed by e.g. electronic transport, spectroscopy, or scanning probe microscopy, their thermal signatures remain largely unexplored. Yet, a full thermal characterization is essential—not only for optimizing heat management in nanoelectronics and designing low-dissipation quantum devices, but also for uncovering novel phases of matter where heat flow and quantum coherence are deeply intertwined.

To address the challenge of characterizing heat transport at the nanoscale, several local thermometry techniques have been developed. These methods can be broadly classified by their sensing mechanism:
(i) Optical techniques, which probe temperature-dependent material properties via light;
(ii) Electron beam-based methods, derived from scanning electron microscopy (SEM), which use beam-induced heating combined with secondary thermometry; and
(iii) Scanning probe techniques, where a nanoscale tip interacts directly with the sample to measure local temperature or thermal conductivity.

Within the optical category, lock-in infrared thermography (LIR) detects black-body radiation emitted by the sample. This approach requires precise calibration of the sample’s emissivity and becomes impractical at cryogenic temperatures due to the steep $T^4$ drop in radiated power below liquid nitrogen temperatures \cite{cifuentes_simultaneous_2017,huth_lock-ir-thermography_2001}. Raman spectroscopy is a widely used technique in solid-state physics that analyzes inelastic light scattering to extract key material properties. Certain Raman peaks are linked to phonon-mediated processes and exhibit temperature dependence, making the method suitable for thermometry. At cryogenic temperatures, reduced thermal broadening enhances sensitivity, although the technique remains material-dependent and typically limited to micrometric spatial resolution. For example, Raman thermometry has been successfully applied to \ch{ZnTe} in the 20–100 \si{\kelvin} range, demonstrating its potential for low-temperature applications.\cite{borisov_znte_2023} Thermoreflectance (TR) is an optical technique that links surface reflectivity to temperature through temperature-dependent optical parameters. A pump laser locally heats the sample, while a probe laser detects the resulting change in reflectivity. TR provides micrometric spatial resolution at room temperature and can be used to extract thermal conductivity. However, its implementation at cryogenic temperatures remains limited—typically above 100\si{\kelvin}—due to experimental complexity, alignment sensitivity, and the need for accurate material parameters such as refractive index and heat capacity across the target temperature range.\cite{wang_characterization_2009,choi_high-speed_2013} Near-field optical thermometry (NFOT), unlike the previous two techniques, overcomes the diffraction limit through near-field interactions, enabling spatial resolution down to 50\si{\nano\meter} \cite{goodson_near-field_1997}. Its fiber-based implementation is inherently compatible with cryogenic environments, and a related technique—scanning near-field optical microscopy (SNOM)—has been demonstrated at temperatures as low as 6\si{\kelvin} \cite{luo_high_2019}. However, to our knowledge, NFOT itself has not yet been experimentally realized under cryogenic conditions.

The second category comprises more specialized electron-based techniques : Scanning Electron Microscopy assisted thermography (tSEM), Electron Thermal Microscopy (ETM) and plasmon transmission electron microscopy (pTEM). In the former, a focused electron beam is used to locally heat up a sample. It relies on a microfabricated thermometer nearby to measure the effective thermal conductance of the system. Since it requires a SEM chamber to operate, it is only compatible with liquid nitrogen cooled sample holders and thus limited to 77 K. \cite{yuan_adapting_2020} The latter relies on the monitoring of metallic nanoparticles near their melting point. It suffers from the same limitation for its operation temperature and requires the coating of the surface with nanoparticles, making it incompatible with numerous sample types. \cite{brintlinger_electron_2008} Both techniques are also poorly suited for insulating samples, as surface charge accumulation disturbs the electron beam. The latter technique, pTEM holds the record for the best spatial resolution, below 1 nm. The technique relies on the monitoring of the temperature dependent energy dispersion created by plasmons and thus is limited to conductive samples. \cite{chmielewski_nanoscale_2020}

The third category includes \textbf{scanning probe techniques}, in which a sharp tip is scanned across the sample to produce spatially resolved maps of temperature and/or thermal conductivity. One example is \textbf{nitrogen-vacancy thermography (tNV)}, which uses a diamond tip doped with a nitrogen-vacancy center as a sensitive quantum thermometer. A green laser excites a temperature-dependent optical transition, whose frequency shift---on the order of tens of \si{\kilo\hertz}---is read out via luminescence. While this method is both sensitive and robust, it has not yet been demonstrated below liquid nitrogen temperatures due to signal degradation~\cite{fukami_all-optical_2019}, and it is incompatible with strong magnetic fields.

An alternative is \textbf{scanning superconducting quantum interference devices (SQUID) on tip (tSOT)}, which exploit the extreme temperature sensitivity of microfabricated SQUIDs, achieving sensitivities around $10^{-6}$\,\si{\kelvin\per\sqrt{\hertz}}. Spatial resolution, determined by the tip-sample contact area, is approximately 400\,\si{\nano\meter}~\cite{wyss_magnetic_2022}. Certain high-$T_c$ designs operate up to 77\,\si{\kelvin}~\cite{black_high-frequency_1995}, and room-temperature magnetometry has been demonstrated using thermally decoupled probes~\cite{dechert_scanning_1999}. However, tSOT thermometry above 4.6\,\si{\kelvin} remains unrealized, despite its theoretical feasibility. In addition, tSOT is incompatible with high magnetic fields and cannot locally heat the sample, preventing its use for measuring heat fluxes or local thermal conductivity.

The final class of scanning probe techniques relies on the direct integration of a thermometer near the tip apex, forming the (sub)family of \textbf{scanning thermal microscopy (SThM)} methods. These are typically categorized based on the thermometer type: \textbf{thermocouple-based (TC-SThM)} or \textbf{thermoresistive (TR-SThM)}. Probes may operate in either DC or AC mode, offering different sensitivities and thermal response characteristics. A comprehensive overview of the various SThM implementations and the factors influencing their spatial resolution is provided in Section~\ref{sec:sthm}.

\begin{figure}
    \centering
    \includegraphics[width=\linewidth]{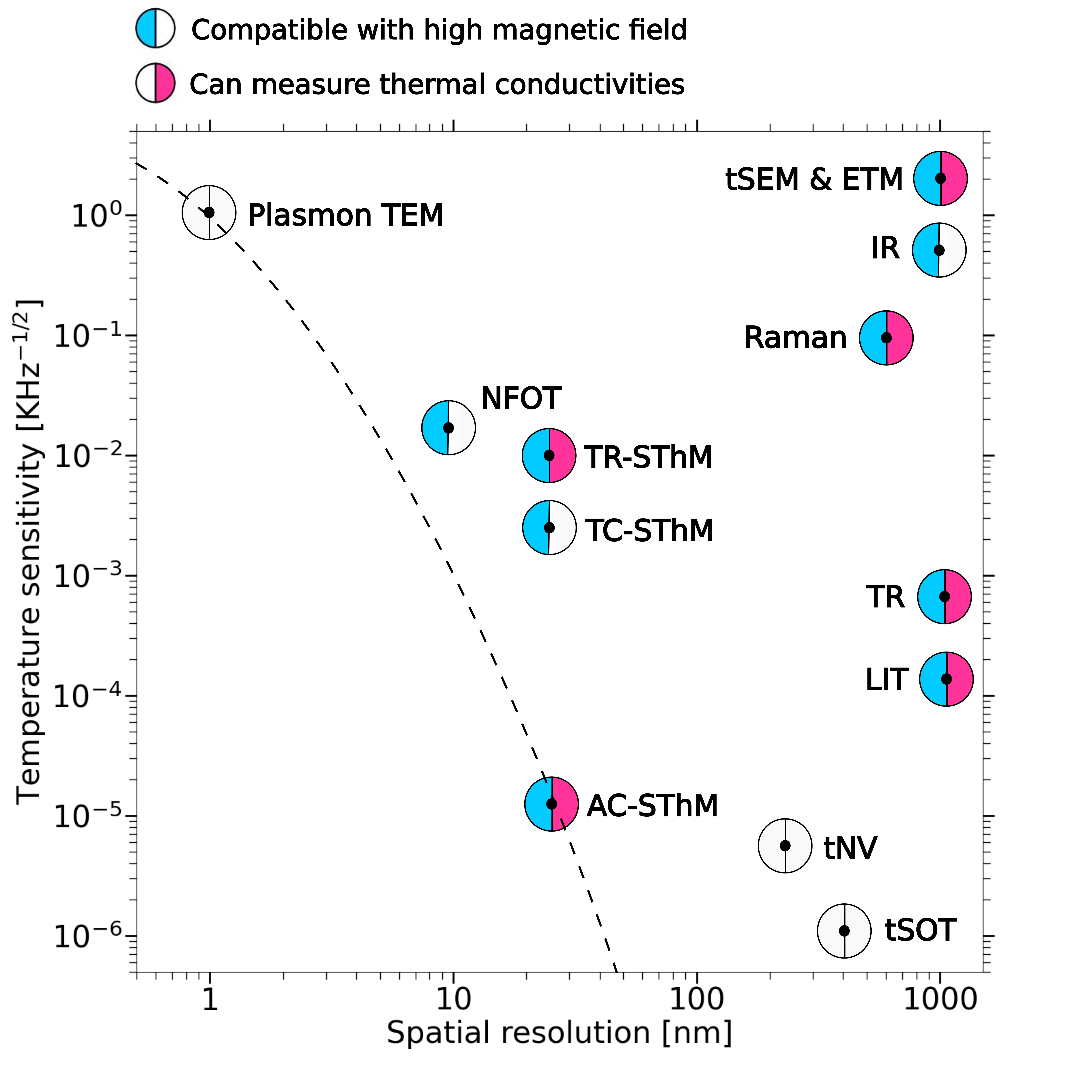}
    \caption{Classification of local thermal characterisation methods by their spatial resolution and temperature sensitivity, the circles surrounding the dot corresponding to a given technique is coloured according to the compatibility of the method with high magnetic field and/or its ability to perform thermal conductivity measurement. The corresponding legend is displayed at the top. The dashed line presents the current frontier in terms of spatial resolution and sensitivity. All the corresponding abbreviations can be found in Section \ref{sec:intro}}
    \label{fig:summary}
\end{figure}

Figure~\ref{fig:summary} compares the \textbf{lateral resolution} and \textbf{temperature sensitivity}—defined as the smallest detectable temperature change per square root bandwidth, in \si{\kelvin\per\sqrt{\hertz}}—across all techniques reviewed. Both metrics are critical for investigating cryogenic solid-state systems, where reduced thermal conductivity and high interface resistance suppress thermal contrast. As a result, extremely high sensitivity becomes essential. As illustrated in the following case studies, many phenomena exhibit nanoscale thermal signatures that currently remain unresolved.

\begin{figure}
    \centering
    \includegraphics[width=\linewidth]{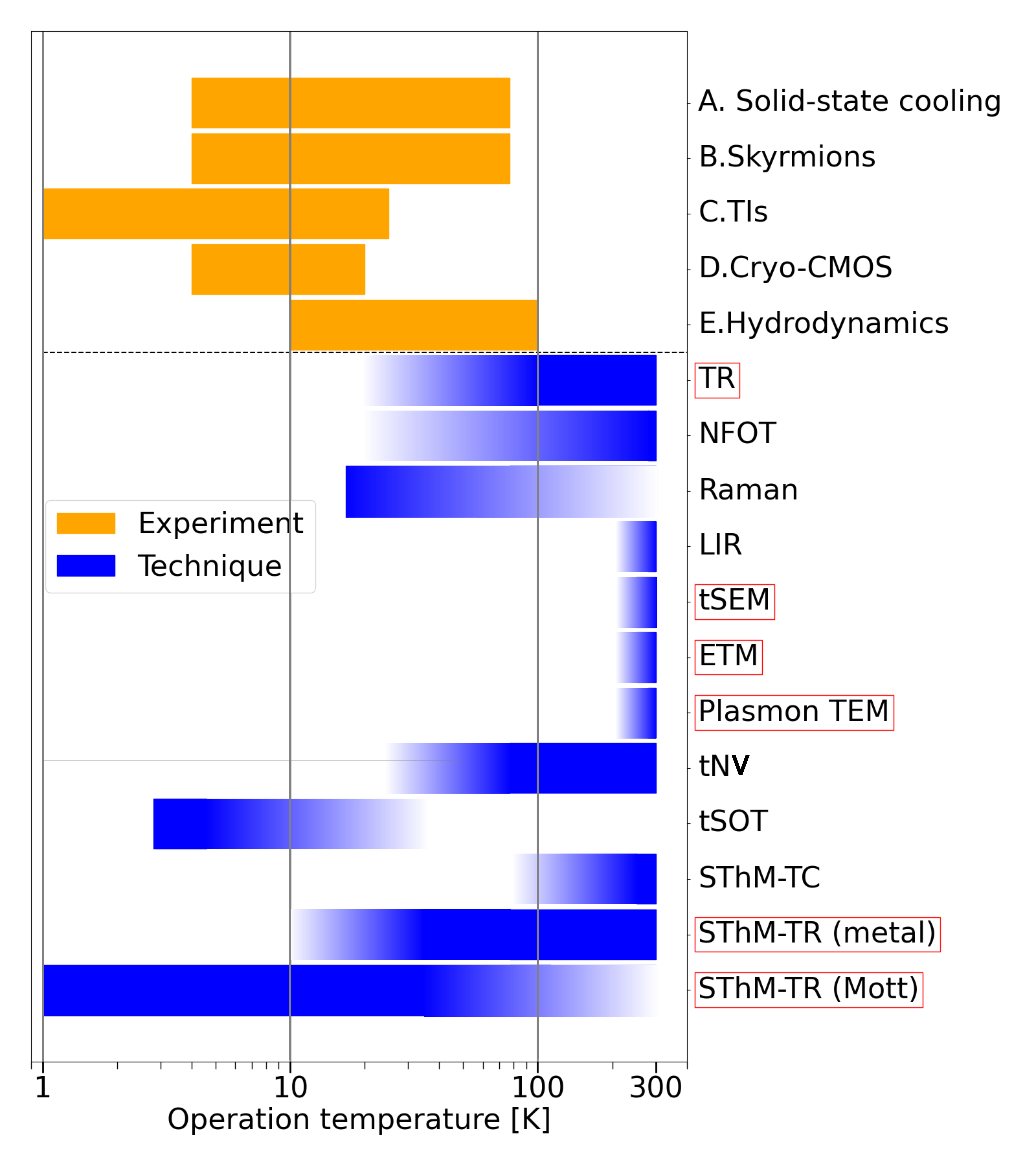}
    \caption{Classification of local thermal characterisation techniques by their window of operation temperature. The hashing of the histogram bar indicates if the thermal characterisation technique has been demonstrated or could be potentially extended at these temperatures. The techniques that are able to measure thermal conductivities are highlighted in red. In orange, the five case studies presented at Section \ref{sec:case_studies} are displayed along with the temperature range where these experiments could be conducted. }
    \label{fig:Tscale}
\end{figure}

Figure~\ref{fig:Tscale} further outlines the \textbf{operational temperature ranges} of each method. Solid bars represent experimentally demonstrated performance, while dashed lines indicate theoretical feasibility that, to our knowledge, has not yet been realized. These gaps typically reflect practical or technological challenges rather than fundamental limitations. The figure also highlights the temperature windows relevant to the case studies discussed later. Notably, while tSOT techniques extend into the ultra-low temperature regime, a significant gap persists between 77\,\si{\kelvin} and 4\,\si{\kelvin}, where no current method enables nanoscale thermal conductivity imaging---representing a key limitation in cryogenic thermal characterization.

\section{Scanning thermal microscopy}
\label{sec:sthm}
Scanning Thermal Microscopy (SThM) is a scanning probe technique where a thermometer is microfabricated close to the apex of an atomic force microscopy (AFM) tip. Since its first realization using a partially exposed Wollaston wire \cite{nonnenmacher_wollaston_1992}, two types of temperature-sensing elements have been used: resistive thermometers (doped semiconductors \cite{nelson_SiProbes_2007} or metals) and thermoelectric junctions, \cite{williams_scanning_1986}. Two distinct operation modes are possible: \textit{passive} or \textit{active} mode. In the former, one parameter (e.g. electrical resistance of the sensing element) is monitored while the sample is heated, while in the latter, the temperature-sensing element is used both as a heater and a thermometer.

\begin{figure}
    \centering
    \includegraphics[width=\linewidth]{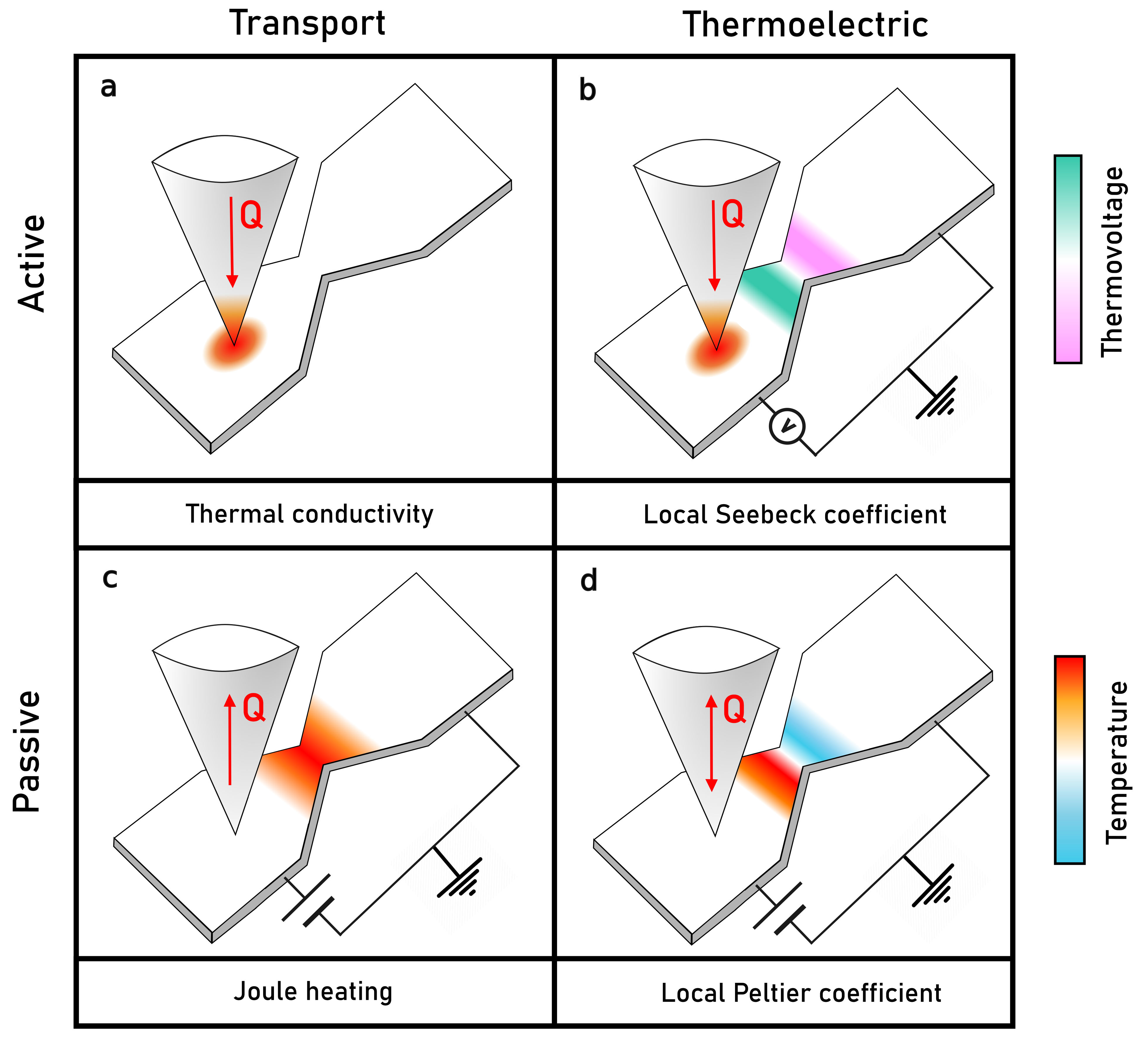}
    \caption{Summary of different SThM modes classified depending on the state of the probe (actively or passively driven) and the properties under study (transport or thermoelectrical) \textbf{a} A hot probe is brought into contact with the surface of the material, the relative change of temperature on the thermometer reflects the thermal conductivity of the surface. \textbf{b} As a hot probe is scanned over the surface, the thermovoltage locally generated  is monitored at the contacts of the sample. This mode is also called Scanning Thermal Gate Microscopy. (STGM) \textbf{c} While the SThM probe is scanned over the surface, a current flows through the sample and generates temperature gradients which are measured by the passively driven SThM probe.  \textbf{d} A current is driven through the sample, if a Peltier junction is present then a thermal gradient will appear which is then detected by monitoring the temperature of the probe.}
    \label{fig:fig1}
\end{figure}

Resistive probes are typically part of a Wheatstone bridge which allows monitoring and quantifying minute changes in the probe resistance. To increase the sensitivity of resistive probes, an AC measurement scheme can be adopted (see Figure \ref{fig:fig1}). The sample/device temperature is modulated, using, for example, an AC bias on the device under test. This creates an oscillating temperature that can be picked-up by the SThM probe using phase-sensitive lock-in detection. Furthermore, this scheme allows to differentiate between heat fluxes originating from different physical effects - e.g. contributions from the Peltier effect and from Joule heating in voltage-biased interconnects - by measuring at different harmonics.

Thermocouple SThM probes rely on the creation of a thermoelectric junction to detect the variation in temperature of a sample. This junction is typically located at the tip apex, making these probes highly sensitive to minute surface temperature variations.

Only resistive probes can operate in an \textit{active} mode, where the thermometer also acts as a nanoscale heat source, as illustrated in Figures~\ref{fig:fig1}a and~b. This unique capability enables the quantification of local thermal \textit{conductivity}, a critical parameter for evaluating properties such as thermoelectric efficiency. Furthermore, active-mode operation provides a dynamic platform to investigate local heat-to-charge conversion phenomena by scanning a heated probe over a device and monitoring its electrical response.

Our group developed such a measurement scheme, termed \textit{Scanning Thermal Gate Microscopy} (SThGM, see Figure~\ref{fig:fig1}b), and demonstrated its utility on systems including graphene devices~\cite{harzheimSThGM2020} and suspended \ch{MoS2} flakes~\cite{razeghi_single-material_2023}. A related, though less quantitative, approach involves using an AFM tip to locally cool a pre-heated sample. While estimating the resulting tip-sample temperature gradient is challenging, this technique has been used for qualitative studies of magnetic materials, including magnetic domain imaging~\cite{budai_high-resolution_2023}.

The primary advantage of SThM lies in its exceptional spatial resolution, which depends on factors such as probe design, operating conditions, and sample properties. Depending on the probe type, resolutions can range from $\sim$10\,nm to several hundred nanometers. Platinum dual-wire resistive probes typically achieve spatial resolutions in this range~\cite{el2017thermal}. Commercially available SThM probes are commonly fabricated from silicon nitride or silicon. Silicon nitride tips usually offer spatial resolution on the order of tens of nanometers~\cite{gomes2015scanning,gonzalez2023direct}, while silicon-based probes can achieve sub-10\,nm resolution under high-vacuum conditions~\cite{menges2016temperature,evangeli2019nanoscale}, which are essential for minimizing air-mediated heat loss. Additional resolution-limiting factors include tip geometry--sharper tips enable finer mapping--contact-mode artifacts, and heat spreading in the sample, especially for highly conductive substrates~\cite{gomes2015scanning}.

As shown in Figure~\ref{fig:Tscale}, implementing a cryogenic-compatible thermometer based on a metal with, for example, a Mott transition—such as \ch{NbN}—could enable SThM operation across the entire temperature range covered in the case studies. Since \ch{NbN} is thermoresistive, it supports all operational modes shown in Figure~\ref{fig:fig1}, offering new avenues for exploring thermal transport and thermoelectric phenomena in exotic quantum materials at ultra-low temperatures. Notably, such thermometers have already demonstrated functionality down to 30\,\si{\milli\kelvin}~\cite{nguyen_niobium_2019}, and similar films have been successfully integrated onto SThM probes by Bourgeois \textit{et al.}~\cite{bourgeois_liquid_2006,swami_experimental_2024}.

Alternatively, thermometric materials such as zirconium nitride (\ch{ZrN}) or zirconium oxynitride (\ch{ZrON}) offer smooth, monotonic resistivity behavior without relying on a phase transition. These materials have demonstrated excellent performance in cryogenic environments down to millikelvin temperatures, owing to their weak electron-phonon coupling, high stability, and compatibility with microfabrication. \cite{yotsuya_new_1987}  Their predictable resistive response and low thermal noise make them attractive candidates for integration into cryogenic SThM probes, particularly where phase transitions (as in \ch{NbN}) may introduce unwanted hysteresis or complexity.

\section{Case studies}
\label{sec:case_studies}
In the following, we present a series of case studies in which the local thermal and/or thermoelectric characterization of exotic phases of matter will transform our understanding of material properties and cryogenic solid-state physics.\\

\subsection*{Case A : Solid-state cooling in the intermediate cryogenic regime}
Thermoelectric cooling at cryogenic temperatures presents both a formidable challenge and a compelling opportunity for next-generation thermal management. The ability to locally regulate the temperature of individual critical components enables substantial energy savings compared to bulk cooling strategies that lower the temperature of entire systems. Beyond efficiency, solid-state cryo-cooling technologies could become transformative in fields such as advanced single-photon detection and quantum computing. For example, they offer the prospect of selectively cooling integrated superconducting qubits below their critical temperature, while maintaining the surrounding CMOS control circuitry at elevated temperatures—-an approach that could drastically enhance both the performance and scalability of quantum architectures.

Among the most promising mechanisms for such targeted cryo-cooling are the Ettingshausen effect (EE) and its anomalous counterpart (AEE)-—the charge-to-heat analogues of the Nernst and anomalous Nernst effects, respectively. In contrast to conventional longitudinal thermoelectric effects, both EE and AEE are \textit{transverse} thermoelectric phenomena, in which the heat flow is orthogonal to the electric current. This transverse configuration enables simpler planar device architectures, reduces thermal losses at interfaces, and alleviates fabrication complexity-—key advantages for integration into compact cryogenic platforms.

In topological materials, the Berry curvature induces a correction to the group velocity in the transport equation.\cite{cooper_Berry_1997,xiao_berry_2005} Xiao \emph{et al.} predicted that this contribution could significantly enhance anomalous transverse thermoelectric coefficients. \cite{xiao_berry_2005} Notably, Weyl semimetals have demonstrated significant Nernst (e.g. WTe${}_2$ : $778\unit{\uV\per\kelvin\per\tesla}$ at $11.3\unit{\kelvin}$ \cite{pan_ultrahigh_2022}, NbSb${}_2$ : $89\unit{\uV\per\kelvin\per\tesla}$ at $21\unit{\kelvin}$ \cite{li_colossal_2022} and ZrTe$_5$ : $146\unit{\uV\per\kelvin\per\tesla}$ at 100$\unit{\kelvin}$\cite{wang_giant_2021}), anomalous Nernst (YbMnBi${}_2$ : $6\unit{\uV\per\kelvin}$ and $1\unit{\uV\per\kelvin}$ at $150\unit{\kelvin}$ and $50 \unit{\kelvin}$ respectively \cite{pan_giant_2022}) and anomalous Ettinghausen (WTe${}_2$ : $0.022\unit{\milli\kelvin\meter\per\ampere}$ at 4.2K\cite{volkl_demonstration_2024}) at cryogenic temperatures, highlighting their potential as efficient, solid-state coolers for quantum computing and other advanced technologies.

\begin{figure}
    \centering
    \includegraphics[width=\linewidth]{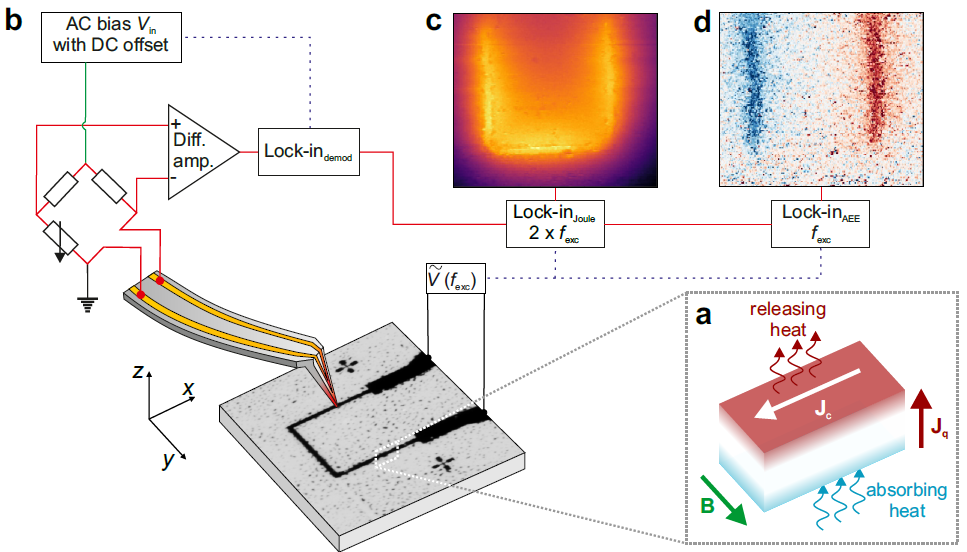}
    \caption{\textbf{a} Illustration of the Ettinghausen effect where a charge current $J_c$ under a magnetic field $B$ creates a perpendicular heat flow $J_q$ in the out-of-plane axis. \textbf{b} Experimental scheme used where the temperature at the apex of a SThM probe is read using a Wheatstone bridge. The probe is scanned over the surface a U-shaped \ch{Co2MnGa} where an AC excitation voltage is applied to its two terminals. \textbf{c} and \textbf{d} are the demodulated SThM signals at respectively the second and first harmonic of the applied voltage, representing the local Joule heating and Anomalous Ettinghausen effect in the device.}
    \label{fig:razeghi_record}
\end{figure}

Our group recently employed thermoresistive SThM probes to locally image the anomalous Ettingshausen effect in \ch{Co2MnGa}, a magnetic Weyl semimetal operating at room temperature.\cite{razeghi_record-high_2024} Temperature variations on the device's upper surface, induced by the Ettingshausen effect, were measured using SThM in an AC excitation scheme,\cite{harzheim_nanocons_2018} as illustrated in Figure~\ref{fig:razeghi_record}. To modulate the local charge carrier mean free path, we used focused ion beam (FIB) etching to create a nanoconstriction-—a strategy previously demonstrated for gold\cite{zolotavin_substantial_2017} and graphene,\cite{harzheim_nanocons_2018} and known to locally influence the Peltier coefficient.

By disentangling the respective contributions from Joule heating and the Ettingshausen effect, and demonstrating sub-100\,nm spatial resolution, this study highlights the power of local SThM characterization for identifying novel \textit{topological} thermoelectric materials with high cooling efficiencies. Furthermore, it opens promising avenues for optimizing nanoscale spot-cooling device designs.

Performing temperature-dependent studies of these behaviours would provide insight into the role of band topology in thermoelectric performance. For this purpose, cryogenic SThM (cryo-SThM) would be an ideal tool, as it directly enables quantitative mapping of spot-cooling power with nanometer-scale resolution.

Ferroelectric materials are a promising class of candidates for cryogenic refrigeration due to their electrocaloric effect, in which entropy changes under the application or removal of an electric field. Under adiabatic conditions, this entropy modulation leads to a reversible heat exchange with the environment, which can be harnessed to create micrometer-scale cooling spots. Although electrocaloric refrigeration has been studied since the 1980s, it is currently experiencing renewed interest as a potential solution to bridge the temperature gap between liquid nitrogen and liquid helium.\cite{radebaugh_electrocaloric_1980,valant_electrocaloric_2012}

Our group recently demonstrated that the localized electric field generated beneath the thermometer element of an SThM probe can be used to image electrocaloric effects with nanometer-scale resolution.\cite{spiece_direct_2025} This technique is compatible with cryogenic operation and offers a powerful tool for nanoscale characterization of polar materials, enabling precise evaluation of electrocaloric performance in microscale refrigeration systems.

\subsection*{Case B : Local thermoelectric effects induced by magnetic textures.}
The exceptional spatial resolution of scanning thermal microscopy (SThM) enables direct, non-invasive detection of thermoelectric signatures arising from individual magnetic textures. This capability has recently allowed the thermoelectric contribution of single domain walls to be resolved in both ferromagnets and antiferromagnets.\cite{puttock_local_2022,isshiki_observation_2024} We anticipate that such techniques can be extended to more complex topological spin textures, including skyrmions and magnetic states in emerging materials such as altermagnets.

While room-temperature magnetic skyrmions have been observed in engineered multilayer thin films over the past decade,\cite{chen_room_2015,yu_skyrmion_2012} their detection remains more robust at cryogenic temperatures and has only been demonstrated under such conditions in two-dimensional magnets.\cite{birch_history-dependent_2022} Thermoelectric imaging of skyrmions has previously been achieved using a UV laser with a $\sim$400\,nm spot size.\cite{iguchi_thermoelectric_2019} However, since skyrmions typically range from a few nanometers to a few tens of nanometers in diameter, the enhanced spatial resolution of SThM could uncover previously inaccessible features — such as interactions with edges, interfaces, and other skyrmions.\cite{brearton_magnetic_2020,capic_skyrmionskyrmion_2020,jackson_skyrmion-skyrmion_1985,arjana_sub-nanoscale_2020} This capability paves the way for a new imaging modality of skyrmion lattices, which are often characterized by reduced skyrmion sizes,\cite{heinze_spontaneous_2011} as well as for studying skyrmions in one-dimensional systems such as nanowires.\cite{wang_thermal_2020} These advancements are particularly relevant for spintronic device applications, where precise skyrmion localization and quantification are critical for technologies like racetrack memories and thermally driven random number generators.\cite{fernandez_scarioni_thermoelectric_2021}  The high spatial and thermal sensitivity of SThM uniquely positions it to accelerate the development and optimization of such spin-caloritronic devices.

Furthermore, the recent discovery of altermagnets -- a new class of magnetic materials that support spin currents along specific crystallographic axes without net magnetization -- introduces a compelling new domain for local thermoelectric imaging.\cite{smejkal_emerging_2022} As their thermoelectric properties are actively investigated, SThM stands out as a key technique for advancing nanoscale understanding and characterization in this emerging field. Embracing these opportunities will allow SThM to push the boundaries of thermoelectric imaging in topological and quantum magnetic systems.
\\

\subsection*{Case C : Quantifying the contribution of edge channels to the Seebeck coefficient of 2D TI}
Disentangling the contributions of edge states and bulk carriers in topological insulators has proven to be a significant challenge. In microscopic samples where only non-local signals are accessible, the bulk typically dominates the overall signal.\cite{xu_enhanced_2014} To our knowledge, no local thermoelectric characterization has yet been performed on 2D topological insulators. By generating a local heat gradient using a heated SThM probe, it is possible to thermally excite charge carriers within the edge states, leveraging the spatial separation between bulk and edge contributions to the material's thermoelectric response. Conversely, applying a current and using a passively driven SThM tip to measure local temperature variations can provide a direct characterization of the Peltier and Ettingshausen effects in these materials — a method that has yet to be explored. As shown in \cite{harzheim_nanocons_2018}, this approach is compatible with a lock-in technique, enabling the separation of Joule and thermoelectric effects based on their distinct frequencies. Such separation is particularly crucial when studying nanoconstrictions, where increased current density can lead to significant local Joule heating.

\begin{figure}
    \centering
    \includegraphics[width=1.0\linewidth]{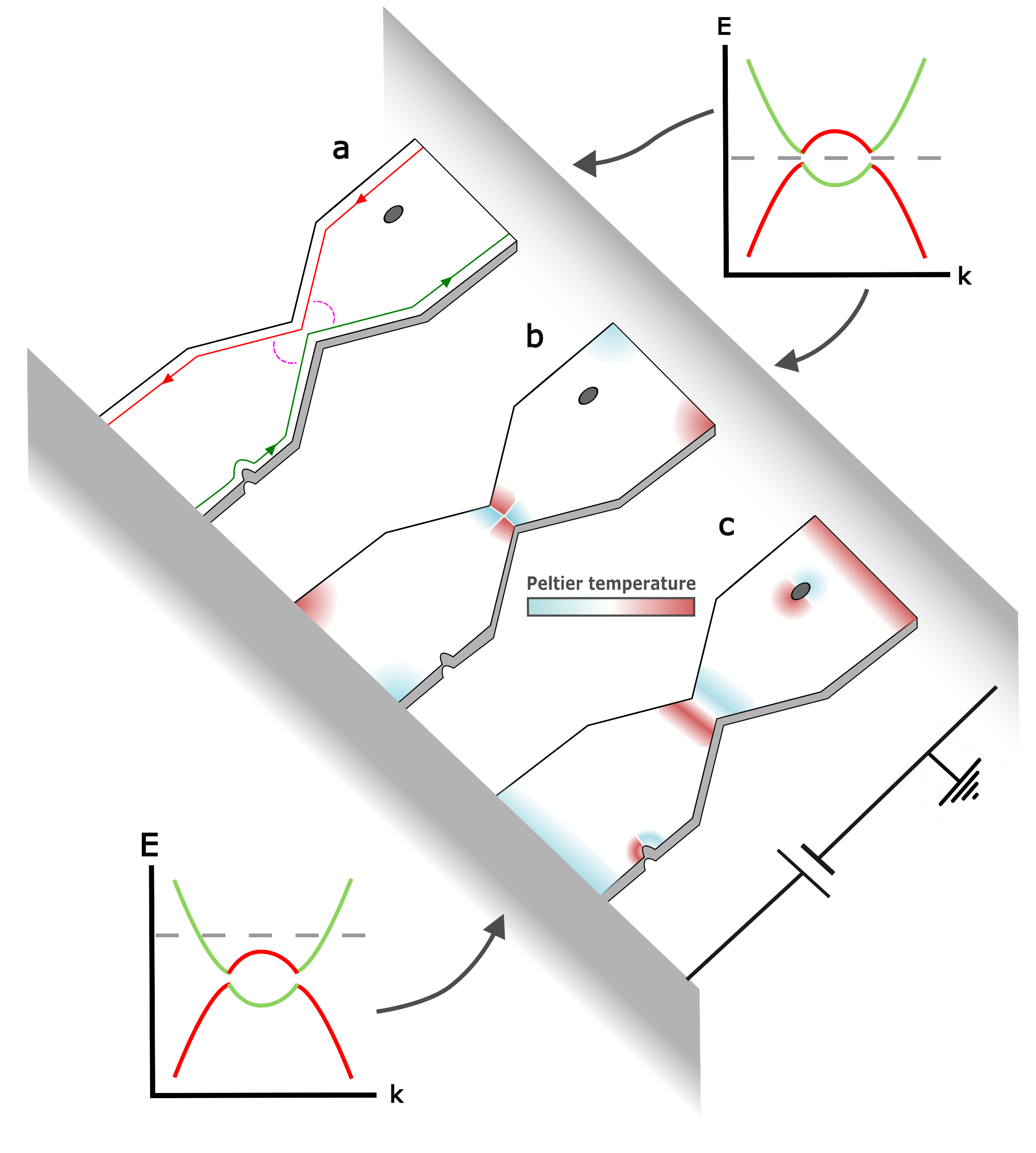}
    \caption{\textbf{a} Edge channels of a 2D topological insulator in the Chern insulator state. \textbf{b} The corresponding temperature map induced by local Peltier effects. A schematic of the material's band structure with his Fermi level as a dashed grey line is presented as an inset. \textbf{c} The temperature map induced by local Peltier effects in the same material as its Fermi level is now shifted in the conduction band.}
    \label{fig:sthm_exp}
\end{figure}

Based on these motivations, we propose the experimental scheme depicted in Figure \ref{fig:sthm_exp}. A bow-tie constriction is etched in a 2D topological insulator, with an AC current applied at frequency $\omega$. When the Fermi level lies within the topological gap, charge transport occurs through topologically protected edge states along the sample boundaries. If the constriction width approaches a few times the localization length, as defined in \cite{xu_enhanced_2014,yang_TITE_Review_2023}, hybridization of the edge channels may occur, introducing a finite probability for backscattering. Due to the topological protection, non-magnetic defects or edge roughness should not disrupt charge flow, leading to no observable contrast in the Seebeck or Peltier signals. In contrast, the presence of a backscattering channel should manifest as a quadrupole-like contrast, indicating variations in the local mean free path of charge carriers. When the Fermi level moves outside the topological gap, the constriction is expected to produce a dipole-like SThM signal, similar to observations in previous studies.\cite{harzheim_nanocons_2018,zolotavin_substantial_2017}

\subsection*{Case D : Thermal management of cryo-CMOS circuitry}

One of the primary challenges in scaling quantum computers lies in the integration of large numbers of qubits. While the fabrication of spin and superconducting qubits has progressed to 300 mm silicon wafers,\cite{elsayed_low_2024} integrating the necessary access and readout electronics remains difficult due to their reliance on conventional CMOS architectures. Cryogenic CMOS (cryo-CMOS) systems are currently under intensive development to address this, aiming to integrate qubit control and readout directly on-chip.

These hybrid systems involve highly dissipative CMOS modules operating in close proximity to thermally sensitive qubits, whose coherence times are strongly affected by local temperature fluctuations. As a result, managing heat dissipation within these systems is critical. Current heat management strategies, however, are largely heuristic due to a lack of local, high-resolution thermal characterization tools -- particularly at cryogenic temperatures. Compounding this, there is limited data on the thermal conductivities of CMOS-compatible materials in the cryogenic regime.

Thermal transport at cryogenic temperatures is often dominated by interface effects, which become increasingly relevant in heterogeneous cryo-CMOS architectures composed of multiple layers and deposition techniques. Accurate characterization of thermal conductivity at the nanoscale is therefore essential to guide chip design.

Scanning thermal microscopy (SThM), especially in cryogenic environments, offers a unique opportunity to address these challenges. It enables the \emph{in situ} mapping of temperature distributions across active cryo-CMOS circuits, while simultaneously probing local thermal conductivities. Importantly, the buried nature of many heat-dissipating components -- often located beneath insulating layers such as \ch{Si3N4}—does not preclude SThM imaging. Techniques such as phase-sensitive detection, analogous to Scanning Thermal Wave Microscopy (STWM),\cite{kwon_scanning_2003} allow probing of thermal propagation speed through overimposed layers. Additionally, strategies such as the null-point method\cite{hwang_quantitative_2016} can be employed to minimize contact thermal resistance, improving measurement accuracy. Moreover, by varying the excitation frequency in an AC-SThM scheme, it is possible to distinguish between heat sources located at different depths or spatial regions. This frequency-domain approach enables selective imaging of adjacent active devices within the same cryo-CMOS chip.

The ability of cryo-SThM to provide both temperature mapping and local thermal conductivity measurements using a single probe offers an unprecedented level of insight into heat flow in quantum-compatible architectures. Such capability is essential not only for validating current heat management strategies but also for guiding the development of new designs that enable the reliable co-integration of qubits and CMOS circuitry at scale.

\subsection*{Case E : Nanoscale heat transport in the hydrodynamic regime}

The hydrodynamic regime of electron transport arises when electron–electron scattering dominates over momentum-relaxing processes such as impurity or boundary scattering. Because these interactions conserve momentum, the electronic system exhibits fluid-like behaviour, reminiscent of classical viscous flows. Experimental signatures of this regime—such as Poiseuille current profiles—have been observed in a range of materials and device geometries.\cite{narozhny_anti-poiseuille_2021,vool_imaging_2021,ku_imaging_2020}

These phenomena typically emerge at intermediate temperatures (10–100\,K), where momentum-conserving electron–electron scattering is still frequent enough to compete with momentum-relaxing processes. Given the temperature sensitivity of this transport regime, cryogenic SThM is particularly well-suited for probing local thermal and electronic behaviour across the relevant temperature range, offering a powerful platform to explore hydrodynamic transport at the nanoscale.

A hallmark of this regime is the breakdown of the Wiedemann–Franz law due to the strong asymmetry between the heat and charge transport relaxation mechanisms, with $L/L_0$ ratio ranging from 0.05 to 20, where $L$ is the measured ratio of thermal and electrical conductivities while $L_0$ is the Sommerfeld value, as demonstrated in numerous studies.\cite{jaoui_departure_2018,zarenia_disorder-enabled_2019, zarenia_thermal_2020, coulter_microscopic_2018, gooth_thermal_2018, lucas_transport_2016, crossno_observation_2016,Huang2024} However, these studies are all spatially averaged, relying on macroscopic samples and integrated microfabricated thermometers. As a result, they cannot resolve how local structural features, such as obstacles or constrictions, affect thermal transport — in contrast to charge transport studies, where local effects have been successfully imaged. \cite{gusev_stokes_2020,alekseev_hydrodynamic_2023,levin_obstacle-induced_2025}

All observed hydrodynamic flows to date have been viscous in nature. A landmark study by Bandurin \emph{et al.} demonstrated the “negative resistance” effect: a contact geometry that induces electron whirlpools flowing counter to the applied voltage, resulting in an apparent negative resistance.\cite{bandurin_negative_2016} This enabled a direct estimate of electronic viscosity ($\sim 0.1\unit{\meter\square\per\second}$), significantly larger than that of honey. While these flows are still laminar, theoretical work has predicted that turbulence could emerge via material engineering, MHz-frequency excitation, or obstacle patterning.\cite{moessner_pulsating_2018} Achieving and imaging turbulent electron flows remains a major experimental goal in solid state physics, with significant implications for energy and information transport. As in classical fluids, turbulence is expected to drastically alter heat transport. Thus, direct thermal imaging at the nanoscale will be essential to fully understand the interplay between turbulence and thermal conductivity. The nanometric spatial resolution of SThM makes it uniquely suited to this task.

\begin{figure}
    \centering
    \includegraphics[width=\linewidth]{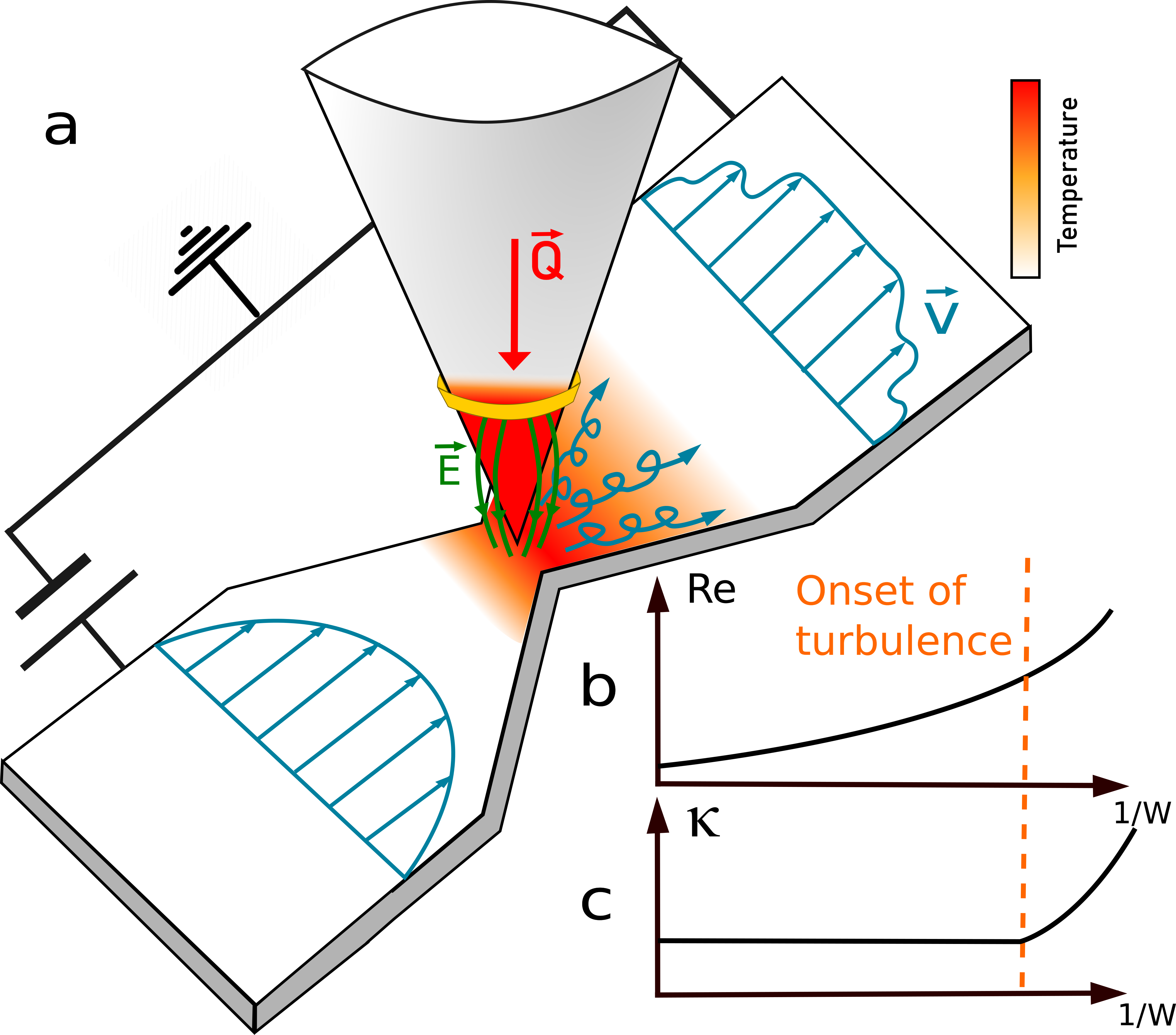}
    \caption{\textbf{a} Schematic experiment to investigate the heat transport in a turbulent, or pre-turbulent, hydrodynamic electron flow. A SThM probe is scanning over a micrometric constriction in active mode. As the generated electric field, in green, perturbs the electron flow in the constriction, the effective width $W$ is reduced, triggering the turbulence of electrons in its wake. This is pictured with the two velocity profiles, in blue, before and after the constriction. As electrons forms whirlpools after the constriction the local thermal conductivity is improved, shown by the temperature gradient in nuances of red. \textbf{b} and \textbf{c} Respectively the Reynolds number $Re$ and the thermal conductivity of the constriction plotted against the inverse of the effective width.}
    \label{fig:hydrodynamic}
\end{figure}

In Figure \ref{fig:hydrodynamic}, we propose an experimental strategy that leverages another key feature of SThM: its capacity to locally apply electric fields. As recently demonstrated in the study of electrocaloric materials,\cite{spiece_direct_2025} thermoresistive SThM probes can induce significant electric fields between their tip and a bottom gate. We propose exploiting this capability in a micrometer-scale constriction to locally perturb the hydrodynamic flow and trigger turbulence. Simultaneously, the probe can measure local variations in thermal conductivity, providing direct insight into how electron whirlpools and turbulent features modify heat transport. This approach opens a new path toward real-space, dynamic visualization of heat flows in the hydrodynamic regime.

\section{Conclusion}

In this Perspective, we have surveyed the most widely used local thermal and thermometry techniques available in the literature. By organizing them according to spatial resolution, temperature sensitivity, operating range, compatibility with thermal conductivity measurements, and resilience to high magnetic fields, we identified a notable gap in local thermal characterization capabilities within the 4-77\si{\kelvin} range. This gap underscores the urgent need for the development of cryogenic scanning thermal microscopy (cryo-SThM).

To provide a comprehensive overview, we reviewed the various thermometer technologies compatible with SThM probes, explored the available operational modes, and outlined the key factors that influence spatial resolution. We then presented five case studies illustrating systems where cryo-SThM could uniquely address current limitations -- whether due to operational temperature constraints, the need for nanoscale thermal conductivity mapping, or local thermoelectric measurements.

By resolving local thermoelectric phenomena at the nanoscale, cryo-SThM not only offers new insights into the thermal transport at cryogenic temperatures but also shines a new light onto previously studied exotic states of matter.  As research in these fields progresses, SThM has the potential to play a transformative role in developing novel materials and device concepts for cryogenic applications.

\section*{Acknowledgements}
The authors acknowledge financial support from the F.R.S.-FNRS of Belgium (FNRS-CQ-1.C044.21-SMARD, FNRS-CDR-J.0068.21-SMARD, FNRS-MIS-F.4523.22-TopoBrain, FNRS-PDR-T.0128.24-ART-MULTI, FNRS-CR-1.B.463.22-MouleFrits, FNRS-FRIA-1.E092.23-TOTEM), from the EU (ERC-StG-10104144-MOUNTAIN), from the Federation Wallonie-Bruxelles through the ARC Grant No. 21/26-116, and from the FWO and FRS-FNRS under the Excellence of Science (EOS) programme (40007563-CONNECT).

\section*{Conflict of Interest}
The authors declare no conflict of interest.

\section*{Bibliography}
\bibliography{mybib}{}
\bibliographystyle{MSP}
\end{document}